\documentclass[a4paper,11pt]{article}
\pdfoutput=1 
\usepackage{jheppub} 
\usepackage[utf8]{inputenc}
\usepackage{amsthm}
\usepackage{mathrsfs} 
\usepackage[all]{xy}
\graphicspath{{./images/}}
\theoremstyle{plain}

\theoremstyle{definition}

\numberwithin{thm}{section}

\newcommand{\vev}[1]{ \left\langle {#1} \right\rangle }

\newcommand{\ket}[1]{ | {#1} \rangle }

\def\d{{\rm d}}
\def\i{{\mathsf i}}

\DeclareMathOperator{\tr}{tr}

\usepackage{slashed}

\def\cA{{\cal A}}

\def\cF{{\cal F}}

\def\cH{{\cal H}}
\def\cI{{\cal I}}

\def\cN{{\cal N}}

\def\cT{{\cal T}}

\def\cZ{{\cal Z}}

\def\bC{{\mathbb C}}

\def\bN{{\mathbb N}}

\def\bR{{\mathbb R}}

\def\bZ{{\mathbb Z}}

\def\sF{{\mathsf F}}
\def\sG{{\mathsf G}}

\def\sc{{\mathsf c}}

\def\so{{\mathsf o}}

\def\su{{\mathsf u}}

\def\U{\mathrm{U}}
\def\SU{\mathrm{SU}}

\def\SO{\mathrm{SO}}

\def\Spin{\mathrm{Spin}}

\def\su{\mathfrak{su}}
\def\o{\mathfrak{o}}
\def\so{\mathfrak{so}}

\def\spin{\mathrm{spin}}
\def\pin{\mathrm{pin}}
\def\beq#1\eeq{\begin{align}#1\end{align}}

\def\h{\widehat}
\def\t{\widetilde}
\def\o{\overline}

\def\fh{\mathfrak h}
\def\ff{\mathfrak f}

\title{General anomaly matching by Goldstone bosons }

\preprint{TU-1108}

\author{Kazuya Yonekura}
\affiliation{Department of Physics, Tohoku University, Sendai 980-8578, Japan}

\abstract{ We describe how Goldstone bosons of spontaneous symmetry breaking $G \to H$ can reproduce anomalies of UV theories under 
the symmetry group $G$ at the nonperturbative level.
This is done by giving a general definition of Wess-Zumino-Witten terms in terms of the invertible field theories in $d+1$ dimensions
which describe the anomalies of $d$-dimensional UV theories. 
The hidden local symmetry $\widehat H$, which is used to describe Goldstone bosons in coset construction $G/H$, plays an important role.
Our definition also naturally leads to generalized $\theta$-angles of the hidden local gauge group $\widehat H$.
We illustrate this point by ${\rm SO}(N_c)$ (or ${\rm Spin}(N_c)$) QCD-like theories in four dimensions. 
} 

\begin{document}

\maketitle

\section{Introduction}
It is commonly believed that spontaneous symmetry breaking is a viable solution to the 't~Hooft anomaly matching condition.
More precise meaning of this statement is as follows. 

Consider a UV theory $\cT_{\rm UV}$ with a global symmetry group $G$. The group $G$ can include spacetime symmetry as well as internal symmetry.
The UV theory is assumed to have an anomaly under $G$. 
Suppose that the symmetry $G$ is spontaneously broken down to a subgroup $H \subset G$ in the IR. 
We assume that the low energy theory consists of Goldstone bosons $U$ and possibly other degrees of freedom which transforms under $H$.
We denote the degrees of freedom other than the Goldstone bosons as $\cT_{\rm IR}$.
The 't~Hooft anomaly matching condition requires that the anomalies of $\cT_{\rm UV}$ and $\cT_{\rm IR}$ under the unbroken symmetry $H$ must be the same.
However, regardless of whether the symmetry is broken or not, the full anomaly must be matched between UV and IR.
The statement made above is that Goldstone bosons can account for ``the rest of the anomaly'' of $G$ which is not produced by $\cT_{\rm IR}$
so that the total IR system $U + \cT_{\rm IR}$ can reproduce the complete anomaly under $G$.

The purpose of this paper is to show that the above expectation is indeed the case. 
At the perturbative level, it is already known that the anomaly matching by Goldstone bosons is possible. 
See \cite{Weinberg:1996kr} for a review of the case relevant to QCD in four dimensions, 
and \cite{Wu:1984pv,Manes:1985df,Hull:1990ms,DHoker:1994rdl,DHoker:1995mfi,Chu:1996fr} for more general cases.
Global anomaly matching is also known in some cases~\cite{Witten:1983tw}. 

We will discuss the general case of any spacetime dimension $d$ with any ``0-form'' symmetry $G$ and any anomaly
at the nonperturbative level.\footnote{However, we exclude the case of conformal anomalies as an exception. 
This case is not (yet) accommodated into the general framework of the description of anomalies by invertible field theories.
Conformal anomaly matching~\cite{Schwimmer:2010za} sometimes has very interesting applications. See e.g. \cite{Komargodski:2011vj}. }
We expect that the argument presented in this paper can also be extended to higher form symmetries~\cite{Gaiotto:2014kfa}. 
In fact, similar ideas have appeared in e.g.~\cite{Kobayashi:2019lep,Hsieh:2020jpj} for higher form anomaly matching 
by higher form analogs of Goldstone bosons in some cases.

The crucial part of our discussion is a definition of general Wess-Zumino-Witten (WZW) terms of Goldstone bosons~\cite{Wess:1971yu,Witten:1983tw}. 
The  present paper is partly motivated by some topological issues of the WZW terms which appear in four 
dimensional QCD-like theories. In \cite{Freed:2006mx,Lee:2020ojw},
it is shown that the WZW terms of four dimensional QCD-like theories with odd color numbers require spin (or $\spin^c$) structure of spacetime. 
This is natural since the UV theories consist of fermions coupled to gauge fields. 
The Atiyah-(Patodi)-Singer index theorem plays an important role for the description of 't~Hooft anomalies of UV fermions,
and it is possible to show the well-definedness of the WZW terms in QCD-like theories by using the index theorem~\cite{Yonekura:2019vyz} which requires spin structure. 

Our point of view in this paper is that whatever structure (e.g. $\spin, \spin^c, \pin^{\pm},$ etc.) which is necessary to define the UV theory, 
or the UV anomaly, can be used to define 
a WZW term in the IR. Given any UV anomaly, we define the corresponding WZW term up to a choice of some generalized $\theta$-angles of Goldstone bosons.
This choice of $\theta$-angles is determined by the UV theory. As an example, we will discuss the case of four dimensional $\SO(N_c)$ or $\Spin(N_c)$
QCD-like theories where there is a nice correspondence between a choice of discrete $\theta$-angles in UV and IR. 

\section{Anomalies and invertible field theories} 
In this section we would like to review the fact that
anomalies of $d$-dimensional theories are described by $(d+1)$-dimensional bulk theories,
which are sometimes called symmetry protected topological phases or invertible field theories.
Here we use the terminology invertible field theory for concreteness. 
The following understanding is formally explained in \cite{Freed:2014iua,Monnier:2019ytc},
and physically concrete arguments are given in \cite{Witten:2015aba,Yonekura:2016wuc,Witten:2019bou} for the case of fermions. 
In the following, $X$ is always a $d$-dimensional manifold and $Y$ is always a $(d+1)$-dimensional manifold. 
They are always equipped with a $G$-bundle and its connection, which is the background fields
of the global symmetry $G$.

An invertible field theory $\cI$ in $d+1$ dimensions is defined by the property that 
its Hilbert space on any closed $d$-manifold $X$ is one dimensional. More physically,
if we consider the theory on $\bR \times X$ where $\bR$ is time and $X$ is space,
the Hilbert space $\cH(X)$ contains a nondegenerate ground state $\ket{\Omega}$
and all other states have a huge mass gap and we neglect these states. 
Thus we can consider $\dim \cH(X)=1$.
Moreover, the partition function $\cZ_\cI (Y)$ on any $(d+1)$-manifold $Y$ 
has absolute value one (if we choose counterterms appropriately), that is $|\cZ_\cI (Y)| =1$.

If we consider the theory $\cI$ on a $(d+1)$-manifold $Y$ with boundary $\partial Y =X$,
the axioms of quantum field theory (or the axioms of path integral when it is available)
say that we get a state vector $\cZ_\cI (Y) \in \cH(X)$ in the Hilbert space $\cH(X)$ of the boundary manifold $X$. 
Notice that $\cH(X)$ is a one-dimensional Hilbert space and hence $\cZ(Y)$ is proportional to the ground state $\ket{\Omega}$.
However, in general there is no canonical choice of the phase of $\ket{\Omega}$ and hence we cannot regard $\cZ(Y)$ as a complex number taking values in $\bC$.
It must be regarded as a vector in $\cH(X)$ even though it is one-dimensional. 

With the above setup, the nonperturbative description of anomalies is as follows. 
A $d$-dimensional anomalous theory $\cT$ whose anomaly is given by the invertible field theory $\cI$ has
the property that its partition function $\cZ_\cT(X)$ on a closed $d$-manifold $X$ is not a complex number, but takes values in $\cH(X)^*$,
i.e. the dual of the Hilbert space $\cH(X)$ of the invertible field theory. 
Notice that nonanomalous theories have partition functions taking values as $\cZ_\cT(X) \in \bC$. 
On the other hand, anomalous theories have partition functions taking values as $\cZ_\cT(X) \in \cH(X)^*$.
If we consider a manifold $Y$ with boundary $\partial Y =X$ and put the theory $\cI$ on $Y$ and 
the theory $\cT$ on $X$, the total partition function is
\beq
\cZ_\cT(X) \cdot \cZ_\cI(Y) \in \cH(X)^* \otimes \cH(X) = \bC,
\eeq
where we have used $\dim \cH(X)=1$ in the equality $ \cH(X)^* \otimes \cH(X) = \bC$.
Therefore, the total bulk-boundary system has the partition function which takes values in $\bC$. 

In other words, if we want to make the partition function $\cZ_\cT(X)$ of the anomalous theory $\cT$ 
to be a well-defined complex number in $\bC$, we must add the bulk $Y$ with the invertible field theory $\cI$ on it. 
In this way, the partition function of the anomalous theory depends on a choice of the bulk $Y$ if we want to get a number in $\bC$.

\begin{figure}
\centering
\includegraphics[width=1.0\textwidth]{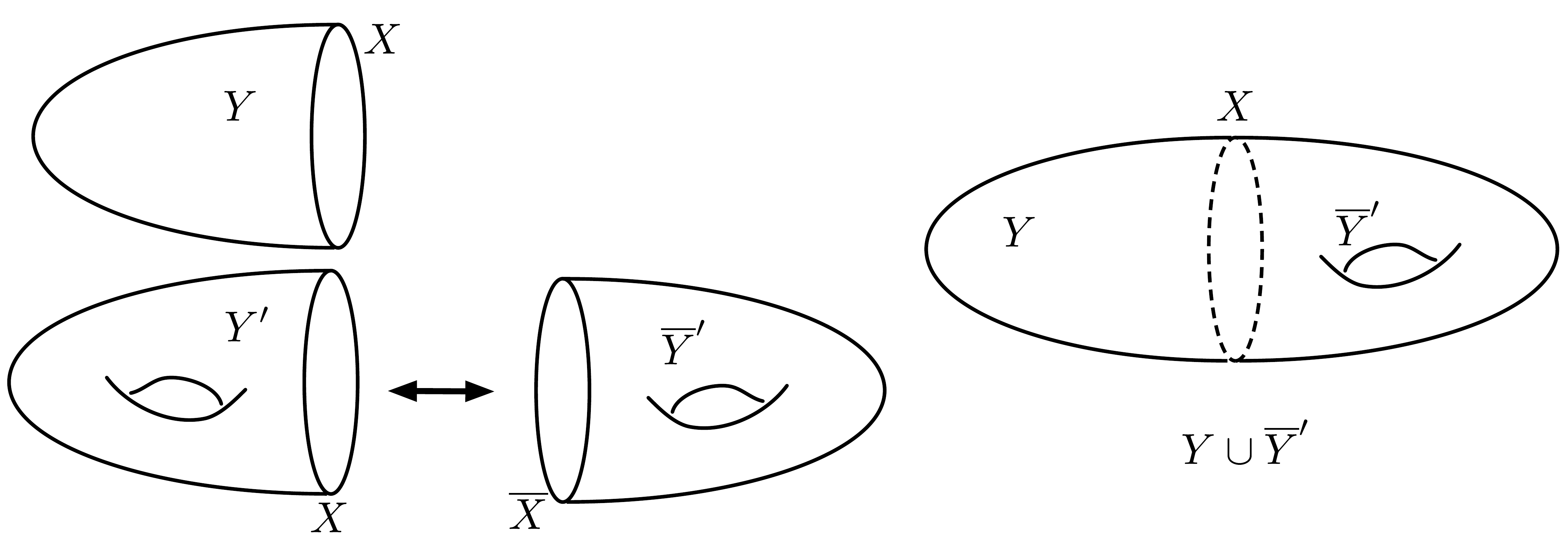}
\caption{ Schematic pictures of manifolds $Y$ and $Y'$ with boundary $X$, the reversal $\o Y'$ of $Y'$, and the glued manifold $Y \cup \o Y'$. \label{fig:1}}
\end{figure}

If we choose a different $(d+1)$-manifold $Y'$, then the difference of the partition function is given by
\beq
\frac{\cZ_\cT(X) \cdot \cZ_\cI(Y)}{\cZ_\cT(X) \cdot \cZ_\cI(Y')} = \cZ_\cI(Y')^* \cZ_\cI(Y)
\eeq
where $\cZ_\cI(Y')^* \in \cH(X)^*$ is the complex conjugation of $\cZ_\cI(Y')$, and we have used $|\cZ_\cI(Y)|=1$. Another axiom of quantum field theory
states the following. Given a manifold $Y$,
there is another manifold $\o Y$ which is roughly the orientation reversal of $Y$ as in Fig.~\ref{fig:1}. 
(But it has a more precise meaning even on non-orientable manifolds; see \cite{Freed:2016rqq,Yonekura:2018ufj}).
The Hilbert space on $X$ and $\o X$ are duals of each other, $\cH(\o X) = \cH(X)^*$.
The quantity $\cZ(Y)$ behaves as $\cZ(Y)^* = \cZ(\o Y)$. By using it, we get
\beq
\cZ_\cI(Y')^* \cZ_\cI(Y) = \cZ_{\cI}(\o Y') \cZ_{\cI}(Y) = \cZ_\cI( Y \cup \o Y'),
\eeq
where $Y \cup \o Y'$ is the closed manifold which is obtained by gluing $Y$ and $\o Y'$ along the boundary $X$.
In the final step of the above equations, we have used the axiom that two ``transition amplitudes'' $\cZ_{\cI}(Y) $ and $\cZ_{\cI}(\o Y') $
can be composed in quantum field theory (or in path integral) along a common boundary. See Fig.~\ref{fig:1}. 
The anomaly is nontrivial if the theory $\cI$ is nontrivial in the sense that $\cZ_\cI(Y_c) \neq 1$ for some closed manifold $Y_c$
($\partial Y_c =0$) even if we choose local counterterms appropriately. See \cite{Witten:2019bou} for more details.

We emphasize that everything is completely gauge invariant in this modern formulation of anomalies. 
Anomalies are not the change of the partition function under gauge variation, but the dependence
of the partition function on the $d+1$-dimensional bulk if we try to define the partition function as a complex number in $\bC$.
More concretely, we can understand anomalous theories as boundary modes of bulk massive theories with
local boundary conditions which preserve gauge invariance. Then we can reach to the above understanding;
see \cite{Witten:2019bou} and Sec.~5.4 of \cite{Hsieh:2020jpj}.

\section{General WZW terms and the anomaly matching}
We consider the situation discussed in the Introduction.
There is a UV theory $\cT_{\rm UV}$ with a global symmetry $G$.
The symmetry is spontaneously broken as $G \to H$, and there are Goldstone bosons $U$
and other IR degrees of freedom $\cT_{\rm IR}$ which transforms under the unbroken symmetry $H$.
We assume that the UV theory has an anomaly which is described by an invertible field theory $\cI$. 
We will define the general WZW term associated to $\cI$ so that the anomalies of UV and IR are matched. 

Before going to the definition of general WZW terms,
we would like to remark that $G$ can contain spacetime symmetries.
For example, we can consider some scalar fields which are odd under parity transformations.
If such parity odd scalars get vacuum expectation values, the parity symmetry is spontaneously broken.
Then we must include the parity symmetry in $G$.
Although the continuous part of the Lorentz group is not spontaneously broken,
there is no problem to include it in $G$. The reason is that 
if $G$ contains the continuous Lorentz group, then we also include it in the unbroken group $H$
so that the coset $G/H$ only contains Lorentz scalars. Therefore, we need not distinguish
spacetime and internal symmetries.
A general framework of 0-form symmetry groups including both spacetime as well as internal symmetries 
is discussed in \cite{Freed:2016rqq} (see also \cite{Yonekura:2018ufj}), and the following discussion is possible in that framework.

We describe Goldstone bosons in the coset $G/H$ as follows.
Locally, the Goldstone field $U$ takes values in $G$ with the ``hidden local symmetry'' $\h H$.
The group $\h H$ can be taken to be the same as $H$, but we put the hat to emphasize that this
is a gauge symmetry rather than a global symmetry. However,
we can also consider more general description in which $\h H$ is some cover of $H$.
In that case, $\h H$ acts on $U$ after the projection $\h H \to H$.
Such a generalization is convenient in some cases such as the one studied in Sec.~\ref{sec:theta}. But for the present section it can be just considered as $H$.

We impose a gauge symmetry $\h H$ acting on $U \in G$ as
\beq
U \mapsto U \h h
\eeq
where $\h h \in \h H$.
The global symmetry $G$ acts on $U$ as 
\beq
U \mapsto g^{-1}U, 
\eeq
where $g\in G$.

Globally, we consider fiber bundles for $G$ and $\h H$ which we denote as $P_G$ and $P_{\h H}$,
respectively. The bundle $P_G$ is for the global symmetry $G$, and we consider its connection $A$ as the background field for $G$.
On the other hand, $P_{\h H}$ is for the gauge symmetry $\h H$. We divide the theory by the gauge transformation group of $P_{\h H}$
so that the physical degrees of freedom of the Goldstone bosons $U$ are given by $G/H$.

From $A$ and $U$, we can construct a connection on $P_{\h H}$ as follows. 
First recall that the background field $A$ transforms under gauge transformation as
\beq
A \mapsto g^{-1} A g + g^{-1} \d g.
\eeq
Then we can define a new gauge field $A_U$ as
\beq
A_U = U^{-1} A U + U^{-1} \d U. \label{eq:AUdef}
\eeq
It is straightforward to check that $A_U$ is invariant under gauge transformations of $G$.
On the other hand, it transforms under gauge transformations of $\h H$ as
\beq
A_U \mapsto \h h^{-1} A_U \h h + \h h^{-1} \d \h h.
\eeq 
Now we decompose the Lie algebra ${\mathfrak g}$ of $G$ as
${\mathfrak g} = {\mathfrak h} \oplus {\mathfrak f}$, where $ {\mathfrak h} $ is the Lie algebra of $H$,
and ${\mathfrak f}$ is a subspace of ${\mathfrak g}$ such that ${\mathfrak g} = {\mathfrak h} \oplus {\mathfrak f}$ holds as a linear space,
and ${\mathfrak f}$ transforms in some representation of $H$.\footnote{
There is some arbitrariness in the choice of ${\mathfrak f}$. 
An inspection of the following discussion of the WZW term suggests that different choices of ${\mathfrak f}$ lead to difference by terms
which are well-defined in $d$ dimensions without using $d+1$ dimensions, and hence not related to anomalies at all. We do not discuss this point in further details. 
}
Under this decomposition, we decompose $A_U$ as
\beq
A_U = A_U^\fh + A_U^\ff,\label{eq:AU}
\eeq
where $A_U^\fh \in \fh$ and $A_U^\ff \in \ff$.
Then $A_U^\fh $ is a connection of the bundle $P_{\h H}$, while $A_U^\ff$ transforms homogeneously (i.e. without the $\h h^{-1} \d \h h$ term)
under gauge transformations of $\h H$. Notice that $A_U$ is nontrivial even if the background field is zero, $A=0$.

Now we can give a definition of the general WZW term $\Phi_{\rm WZW}$ associated to the invertible field theory $\cI$ in $d+1$ dimensions. 
The anomalous theory $\cT_{\rm UV}$ is living on a $d$-manifold $X$. 

First, we take a $(d+1)$-manifold $Y$ with $\partial Y = X$.
The background field $A$ is extended to $Y$. We assume that such a $Y$ exists, and will comment on 
the case that such a $Y$ does not exist for a given $X$ in Sec.~\ref{sec:theta}. 

Second, we consider a manifold $[0,1] \times X$.
We denote the coordinate of the interval $[0,1]$ as $s$, and take a function $\rho(s)$
which interpolates between $\rho(0)=1$ and $\rho(1)=0$ smoothly. It is possible to show that the following construction does not depend on how to take $\rho(s)$.
Then we consider a gauge field on $[0,1] \times X$ as
\beq
A_U^\fh + \rho(s)A_U^\ff \label{eq:interpolate}
\eeq
where $A_U^\fh $ and $A_U^\ff $ are defined as above in terms of the Goldstone field $U$ and the gauge field $A$ on $X$.
Notice that this gauge field is $A_U$ at $s=0$ and $A_U^\fh$ at $s=1$.

Third, we take a manifold $ Y_0 $ whose boundary is $X$ such that the gauge field $A_U^\fh $ on $X$
is extended to $Y_0$ as a $\h H$ gauge field. In Sec.~\ref{sec:theta} we will discuss the case that such a $ Y_0$ does not exist,
and here we simply assume the existence of $Y_0$.

\begin{figure}
\centering
\includegraphics[width=0.8\textwidth]{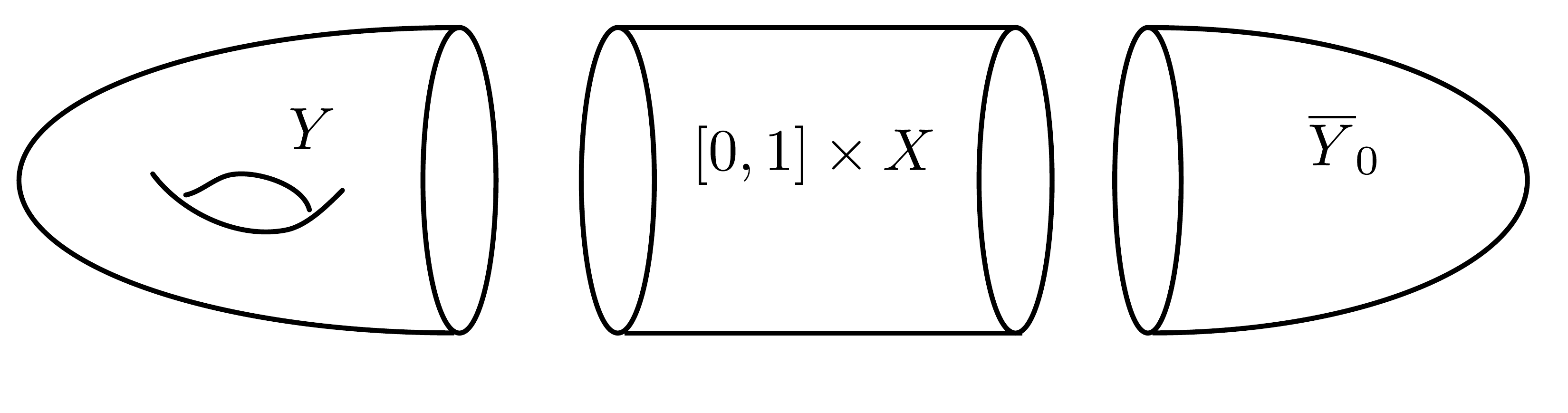}
\caption{ Gluing three manifolds $Y$, $[0,1] \times X$ and $\o Y_0$ to get a closed manifold $Y_{\rm total}$. \label{fig:2}}
\end{figure}

Finally, we glue the three manifolds $Y$, $[0,1] \times X$, and $\o Y_0$ to get
\beq
Y_{\rm total} = Y \cup ([0,1] \times X) \cup \o Y_0.
\eeq
See Fig.~\ref{fig:2}. 
In the gluing between $Y$ and $[0,1] \times X$, we need to use $U$ on $X$ as a transition function.
The reason is that the gauge field on $Y$ is $A$, while the gauge field on the boundary $\{0\} \times X$
of $[0,1] \times X$ is $A_U$, and they are related as in \eqref{eq:AUdef}.
Thus we use $U$ as a transition function between them. 

We define the general WZW term $\Phi_{\rm WZW}$ associated to the invertible field theory $\cI$ as
\beq
\Phi_{\rm WZW} : = \cZ_{\cI}(Y_{\rm total}) \in \U(1).
\eeq
where the right hand side is the partition function of the invertible field theory $\cI$ on $Y_{\rm total}$
with the gauge field configuration specified above. 

Before revealing the properties of the WZW term, we must recall that the IR theory contains the degrees of freedom
$\cT_{\rm IR}$. We define the partition function of this theory as
follows. By the 't~Hooft anomaly matching condition, we must assume that this theory
has the anomaly under $H$ which is the restriction of the invertible theory $\cI$ to only $H$-bundles. 
Then we define the IR partition function (for a fixed configuration of the Goldstone field $U$) by using the above $Y_0$ as
\beq
\cZ_{\rm IR} : = \cZ_{\cT_{\rm IR}}(X)\cZ_{\cI}(Y_0) \in \bC,
\eeq
where $\cZ_{\cT_{\rm IR}}(X) \in \cH(X)^*$ is the partition function of $\cT_{\rm IR}$ which is coupled to $A_U^\fh$.
Notice that this construction requires the coupling of the Goldstone bosons and $\cT_{\rm IR}$
via the gauge field $A_U^\fh$. As remarked before, it is nonzero even if $A=0$, and therefore it is a genuine 
interaction between the Goldstone bosons and the other IR degrees of freedom $\cT_{\rm IR}$. 
Notice also that that we have used the same $Y_0$ as in the definition of $\Phi_{\rm WZW}$.

Now let us look at the important properties of the above quantities $\Phi_{\rm WZW} $ and $\cZ_{\rm IR} $.
\begin{enumerate}
\item Each of $\Phi_{\rm WZW} $ and $\cZ_{\rm IR}$ may depend on $Y_0$. However, their product $\Phi_{\rm WZW} \cZ_{\rm IR}$
is independent of a choice of $Y_0$. The reason is as follows. The axioms of quantum field theory say that 
$\cZ_{\cI}(Y_{\rm total})=\cZ_{\cI}(Y)\cdot \cZ_{\cI}([0,1]\times X)\cdot \cZ_{\cI}(\o Y_0)$, 
where each factor takes values in some one-dimensional Hilbert spaces $\cH(\bullet )$.
In the product $\Phi_{\rm WZW} \cZ_{\rm IR}$, the manifold $Y_0$ appears as $\cZ_{\cI} (\o Y_0) \cZ_{\cI} (Y_0) = |\cZ_{\cI} ( Y_0) |^2=1  $.
Therefore, the $Y_0$ dependence cancels out between $\Phi_{\rm WZW} $ and $\cZ_{\rm IR} $.
A corollary is that if the IR theory $\cT_{\rm IR}$ is empty, the WZW term $\Phi_{\rm WZW} $ is independent of $Y_0$.

\item In our construction, we have used $U$ which is defined only on $X$. Therefore, it is manifest that
the WZW term (or more precisely the product $\Phi_{\rm WZW} \cZ_{\rm IR}$ in which the $Y_0$ dependence cancels out) depends only on 
the $d$-dimensional configuration of $U$. This is somewhat surprising given that the usual construction of WZW terms
requires $U$ to be extended to a higher dimensional manifold, which is $Y$ in the present case. 
The reason we can avoid extending $U$ to $Y$ is that we have used $U$ as a transition function
between $Y$ and $[0,1] \times X$ so that there is no $U$ dependence on $Y$. 

\item The WZW term $\Phi_{\rm WZW}$ depends on $Y$ and the background field $A$ on it.
This is completely the same dependence as the UV theory, and represents the anomaly of the theory. Therefore,
we have established that the IR theory $U + \cT_{\rm IR}$ can reproduce the same anomaly as the UV theory. 

\end{enumerate}
After defining the WZW term, the final IR partition function $\cZ_\text{total~IR}$ is obtained by performing the path integral over $U$ as
\beq
\cZ_\text{total~IR} = \int [D U]  \cZ_{\rm IR} \Phi_{\rm WZW}e^{-S(U)}
\eeq
where $S(U)$ is an action of $U$ which does not contribute to the anomaly. 
This path integral is completely gauge invariant, and this partition function is a functional of the background gauge field $A$.
The anomaly is given by the bulk theory partition function $ \cZ_{\cI}(Y)$.
This concludes the general discussion of the WZW term.

\paragraph{Examples.}
One might feel that the above construction is too abstract.
So let us study a concrete example.
As a UV theory, we consider $d=2n$-dimensional $N_f$ massless Dirac fermions which contain positive chirality components $\psi^i$ and negative chirality components $\t \psi_{\tilde i}$,
where $i, \t i =1, \cdots, N_f$. There are $\U(N_f)_L $ and $ \U(N_f)_R$ symmetries acting on the indices of $\psi^i$ and $\t \psi_{\tilde i}$, respectively. 
We also introduce scalar fields $S^{\tilde i}_{~j}$ in the bifundamental representation of $\U(N_f)_L \times \U(N_f)_R$ and add a coupling 
$S^{\tilde i}_{~j} \psi^j \t \psi_{\tilde i}$.
We assume that $S^{\tilde i}_{~j}$ gets a vacuum expectation value $\vev{S^{\tilde i}_{~j}} \propto \delta^{\tilde i}_{~j}$ 
such that the symmetry is broken to the diagonal subgroup $\U(N_f)_L \times \U(N_f)_R \to \U(N_f)_{\rm diag}$. All the fermions become massive and disappears from the low energy theory. Thus $\cT_{\rm IR}$ is empty.

Just for simplicity, let us neglect $\U(N_f)_R$ and consider only $G = \U(N_f)_L$ which is spontaneously broken to $H =\{1\}$.
We denote $\U(N_f)_L$ just as $\U(N_f)$, and denote the background gauge field for $\U(N_f)$ as $A$. 
The general discussions given above can include both spacetime and internal symmetries in $G$, 
but here we take $G$ to be just (the subgroup of) the internal symmetry, $A$ to be a background field for it,
and treat gravity as different fields. 

We want to obtain a differential form formula for the WZW term.
To simplify the situation further, we assume that all fields (except possibly for gravity) are topologically trivial. 
However, we must recall that we have used $U$ as a transition function between $Y$ and $[0,1] \times X$.
To trivialize the $G$-bundle on the entire space $Y_{\rm total}$, we extend $U$ to $Y$ in an arbitrary way.
Then we use $A_U$ instead of $A$ inside $Y$. In this way the $G$-bundle $P_G$ is trivialized. 
We denote the gauge field constructed on $Y_{\rm total}$
as above as $\cA$.

In this topologically trivial case, the partition function of the invertible field theory of the current model is just given by the Chern-Simons invariant,
\beq
\cZ_{\cI}(Y_{\rm total}) = \exp\left(2\pi \i \int_{Y_{\rm total}} \h {A}(R) I(\cA) \right),
\eeq
where $\h {A}(R) = 1 +\frac{1}{48(2\pi)^2}\tr R^2 + \cdots$ is some invariant polynomial of the Riemann curvature 2-form $R$, and
$I(\cA)$ is given as follows. We introduce a new coordinate $t \in [0,1]$ and consider a $(d+2)$-dimensional gauge field $\t \cA = t \cA$.
Let $\t \cF = \t \d \t \cA + \t \cA^2$ be its field strength in $d+2$ dimensions, where $\t \d = \d t\, \partial_t + \d$ 
is the exterior derivative in $d+2$ dimensions. Then we define
\beq
I(\cA) = \int_{t\in [0,1]} \tr \exp \left(\frac{\i}{2\pi} \t \cF \right), \label{eq:CSform0}
\eeq
where the integral is over $t \in [0,1]$, and the trace is over $\U(N_f)$ indices. 
It is defined so that $I(\cA)$ satisfies $\d I(\cA) = \tr \exp \left( \frac{\i}{2\pi} \cF \right)$, where $\cF = \d \cA + \cA^2$ is the curvature 2-form
in $d+1$ dimensions. A more explicit expression is given by
\beq
I(\cA)=   \int^1_0 \d t\,  \tr \left( \frac{\i}{2\pi} \cA \exp \left[\frac{\i}{2\pi}  (t\, \d \cA + t^2 \cA^2)  \right] \right). \label{eq:CSform}
\eeq
This is the standard Chern-Simons form.

Now let us evaluate the above partition function of the invertible field theory on $Y_{\rm total}$. 
We would like to show that the contribution from the region $[0,1] \times X$ is zero. It can be seen as follows. 
From \eqref{eq:interpolate} applied to the current case $G=\U(N_f)$ and $H=\{1\}$, the gauge field $\t \cA$ in this region is given by
\beq
\t \cA = t \rho(s)  A_U.
\eeq
Notice that the coordinates $s$ and $t$ appear only in the combination $u=t\rho(s)$, since $A_U$ is independent of them.
Then we see that
\beq
\int_{(s,t) \in [0,1] \times [0,1]} \tr \exp \left(\frac{\i}{2\pi} \t \cF \right) =0.
\eeq
The reason is that we can define a new coordinate system $(u,t)$ instead of $(s,t)$, and then
only $\d u $ appears in the integrand without $\d t$. 
Therefore, we conclude that the contribution from the region $[0,1] \times X$ is zero in the current situation.

The gauge field $\cA$ on $Y$ is given by $A_U$ as mentioned above. Therefore, we finally get
\beq
\Phi_{\rm WZW}(U,A) = \cZ_{\cI}(Y_{\rm total}) = \exp\left(2\pi \i \int_{Y} \h {A}(R) I(A_U) \right).
\eeq
A corollary is that if the background field $A$ is zero, the WZW term is given by
\beq
\Phi_{\rm WZW}(U) =  \exp\left(2\pi \i \int_{Y} \h {A}(R) I(U^{-1} \d U) \right).\label{eq:UNfWZW}
\eeq
This is the standard WZW term for the $\U(N_f)$ matrix valued field $U$.

The UV theory, and in particular the invertible field theory of the current model, is well-defined if manifolds have spin structure
because we can define the invertible field theory in terms of the Atiyah-Patodi-Singer $\eta$-invariant for fermions~\cite{Witten:2015aba,Yonekura:2016wuc,Witten:2019bou}.\footnote{
The Atiyah-Patodi-Singer index theorem~\cite{Atiyah:1975jf} (see also \cite{Fukaya:2017tsq,Dabholkar:2019nnc,Fukaya:2019qlf}) states that \eqref{eq:CSform0} integrated over $Y_{\rm total}$
modulo integers is given by the difference 
of the $\eta$-invariants of the Dirac operators which are coupled and not coupled to $A$, respectively.
With our simplifying assumption of neglecting $\U(N_f)_R$, this difference is precisely the UV anomaly. 
} 
Therefore, our general construction implies that the above WZW term $\Phi_{\rm WZW}(U) $ is well-defined with spin structure.
This is proved in \cite{Freed:2006mx,Lee:2020ojw} in the case $d=4$ by using techniques of algebraic topology. 
The argument presented in this paper is a generalization of the one given in \cite{Yonekura:2019vyz}.
Equivalently, the well-definedness of \eqref{eq:UNfWZW} is a consequence of $K^{-1}$ theory~\cite{Lee:2020ojw}.

In the above example, there are no massless fermions in the IR.
To get some insight into the case that there remain massless fermions which are 
charged under $H$, let us consider the following example.

We consider a UV model in which there are $N_f$ left-handed fermions $\psi_i$ and only a single right-handed fermion $\t \psi$.
We introduce a scalar $S^{i} $ in the fundamental representation of $G=\U(N_f)$ which is coupled to the fermions
as $S^i \psi_i \t \psi$.
We assume that the scalar gets a vacuum expectation value
$\vev{S^{1}} \neq 0$ and  $\vev{S^{j}} =0$  for $j \geq 2$.
The symmetry is broken down to $H=\U(N_f-1)$.

We introduce the Goldstone field $U^i_{~j}$
as $S^i = U^i_{~j} \vev{S^j} $, where $U \in \U(N_f)$ and we have neglected the radial direction of $S$.
The scalar $S^i$ is unchanged under $U^i_{~j} \to U^i_{~k} \h h^k_{~j}$ if $\h h \in \U(N_f-1)$,
so this is a new gauge symmetry (or gauge redundancy) which we have called the hidden local symmetry.

Let us change the fermion variables from $\psi_i$ to $\chi_i$ which is defined as
\beq
\chi_i = \psi_j U^j_{~i}. \label{eq:cov}
\eeq
Then the coupling becomes
\beq
S^i \psi_i \t \psi = \vev{S^i}  \chi_i \t \psi = \vev{S^1} \chi_1 \t \psi.
\eeq
The pair $(\chi_1, \t \psi)$ gets a mass term, while other $\chi_i $ for $i \geq 2$ are massless.
The kinetic term of $\chi_i$  is proportional to
\beq
{\psi}_i \gamma^\mu \left( \delta^i_{~j} \partial_\mu + (A_\mu)^i_{~j} \right) \o \psi^{j} 
= {\chi}_i \gamma^\mu \left(  \delta^i_{~j}  \partial_\mu + (U^{-1}A_\mu U+U^{-1} \partial_\mu U)^i_{~j}  \right)  \o \chi^{j} .
\eeq
Therefore, in the low energy, the massless fermions $\chi_i~(i\geq 2)$ interact with the Goldstone field
via the gauge field 
\beq
A_U^\fh : = \left(  (U^{-1}AU + U^{-1} \d U)^i_{~j} \right)_{2 \leq i,j \leq N_f}.
\eeq
This interaction is exactly what we needed to assume in order to show the anomaly matching in the general discussion.
In this calculable example, the interaction with $A_U^\fh $ arises in the above way.

The change of variables \eqref{eq:cov} produces a new contribution to the effective action by the chiral anomaly of $\U(N_f)$.
This contribution roughly corresponds to the part $\cZ_{\cI}(Y_0)$ in the general discussion.
But the precise details are involved and we will not study it further.

\section{Topological theta angles}\label{sec:theta}
In this section we would like to explain some topological issues which are not explained in the previous section.

The first point is about the topology of the background field $A$.
We have seen that the Goldstone field $U$ can be used to change this field to another field $A_U = U^{-1}AU + U^{-1} \d U$
by using $U$ as a gauge transformation. The new field $A_U$ transforms only under the gauge group $\h H$. 
Topologically, this means that the $G$-bundle $P_G$ can be reduced to the $\h H$-bundle $P_{\h H}$.
Then, it is natural to wonder what happens if the topology of the background $P_G$ is such that it cannot be reduced to an $\h H$-bundle.

We claim that we should take the partition function of the Goldstone bosons to be zero (or at least very small in the sense explained below) in the case in which the $G$-bundle is not reduce to an $\h H$-bundle.
To better understand the situation in a concrete example, let us consider the fermion theory discussed in the previous section with the symmetry $\U(N_f)_L \times \U(N_f)_R$
which is coupled to $S = (S^{\tilde i}_{~j} )$. Suppose that the bundles of $\U(N_f)_L $ and $ \U(N_f)_R$ are topologically different.
Then $\det S$ must vanish somewhere in the spacetime. The reason is that if $\det S$ were everywhere nonzero,
$S$ would give a bundle isomorphism between the $\U(N_f)_L $ and $ \U(N_f)_R$ bundles, contradicting with the assumption. 
Now notice that the scalar $S$ may have a potential energy so that the symmetry is spontaneously broken as $\U(N_f)_L \times \U(N_f)_R \to \U(N_f)_{\rm diag}$. 
Then, the points where $\det S$ vanish have large potential energies and they are outside the description of the low energy effective field theory of Goldstone bosons.
The large potential energies make the partition function exponentially small in the low energy limit. 
If the bundles of $\U(N_f)_L $ and $ \U(N_f)_R$ are topologically the same, they can be reduced to the diagonal bundle $\U(N_f)_{\rm diag}$.
This is the reason for the above claim. However, we remark that those ``topological solitons'' in which some fields of the UV theory (such as the $S$ above) go outside
the effective field theory of Goldstone bosons have interesting anomalies. See e.g. \cite{Cordova:2019wpi,Hason:2019akw} for a spontaneously broken $\bZ_2$ as an extreme example
where there are no continuous Goldstone bosons (i.e. $G/H$ consists of discrete points). 

Our next task is to understand the case that the $G$-bundle and the $\h H$-bundle on a $d$-manifold $X$ cannot be extended
to some $(d+1)$-manifolds $Y$ and $Y_0$ which are used to construct the general WZW term in the previous section.
The absence of such extensions implies the possibility of topological $\theta$-angles of the groups.
For example, we can consider a four dimensional sphere $X=S^4$ with a nontrivial instanton of $\U(N_f)$.
The instanton number is given by $N= \int_X \frac{1}{2} \tr \left(\frac{\i}{2\pi} F \right)^2 $.
Suppose that $X=S^4$ is a boundary of some $Y$ on which the $\U(N_f)$-bundle is extended.
Then the instanton number is zero since the Stokes theorem implies $N= \int_X \frac{1}{2} \tr \left(\frac{\i}{2\pi} F \right)^2 = \int_Y \frac{1}{2} \d \tr \left(\frac{\i}{2\pi} F \right)^2 =0$.
If there is a nontrivial instanton on $X$, we cannot extend it to $Y$. In this case we choose the phase of the partition function ``by hand'',
but in a systematic way sketched below.
This choice ``by hand'' corresponds to a choice of the $\theta$-angle $\exp( \i \theta N)$. 
The correspondence between the non-existence of $X$ to $Y$ and $\theta$-angles in a generalized sense is not restricted to the above example,
but is a general fact~\cite{Yonekura:2018ufj}. 

The general procedure is as follows~\cite{Witten:2016cio,Witten:2019bou}. 
Here we explain it for the group $G$, and we assume that all symmetries (including spacetime symmetries) are included in $G$.
First we define the bordism group $\Omega_d^G$ as follows. 
If two $d$-manifolds with $G$-bundles $X_1$ and $X_2$ are realized as a boundary of a $(d+1)$-manifold $Y$ with $G$-bundle
as $\partial Y = X_1 \sqcup \o X_2$ where $\sqcup$ means disjoint union, we regard $X_1$ and $X_2$ to be equivalent, $X_1 \sim X_2$.
This forms an equivalence relation. As a set, $\Omega_d^G$ is a set of equivalence classes, 
and we denote the equivalence class containing $X$ as $[X]$ (i.e., if $X_1 \sim X_2$ then $[X_1] = [X_2]$). 
We can introduce an abelian group structure on $\Omega_d^G$ by $[X_1] + [X_2] = [X_1 \sqcup X_2]\,$, $ [\varnothing]=0\,$ and $ [\o X] = - [X]$.
 In this way the bordism group $\Omega^G_d$ is defined. We assume that the abelian group $\Omega^G_d$ is finitely generated, which implies
 that it is a direct sum of factors which are isomorphic to a group of the form $\bZ$ or $\bZ_n$ for some $n \in \bN$. Namely, $\Omega^G_d$ is of the form
 $\bZ \oplus \cdots \oplus \bZ_{n} \oplus \cdots$.
 
 For a free factor $\bZ$ in $\Omega^{G}_d$, we take a generator $[X^{\rm (ref)}_{\rm free}]$ for it, and define the phase of the partition function 
 on a fixed reference manifold $X_{\rm free}^{\rm (ref)}$ in an arbitrary way.
 This choice of the phase corresponds to a continuous $\theta$-angle.
 For a torsion factor $\bZ_n$ in $\Omega^{G}_d$, we take a generator $[X^{\rm (ref)}_{\rm tor}]$ for it and a fixed reference manifold $X^{\rm (ref)}_{\rm tor}$. 
 We know that $n$ copies of $X^{\rm (ref)}_{\rm tor}$ is a boundary of some $(d+1)$-manifold
 $Y^{\rm (ref)}_{\rm tor}$, and hence the partition function on the $n$ copies of $X^{\rm (ref)}_{\rm tor}$ is defined if we fix this $Y^{\rm (ref)}_{\rm tor}$.
 In this way the product of $n$ copies of the partition function on $X^{\rm (ref)}_{\rm tor}$ is defined for a fixed  $Y^{\rm (ref)}_{\rm tor}$.
 Now we take the $n$-th root of it to get a partition function on a single $X^{\rm (ref)}_{\rm tor}$. 
 A choice of this $n$-th root corresponds to a discrete $\theta$-angle. 
 For an arbitrary $X$, there is a manifold $Y$ whose boundary consists of the $X$ and some copies of the reference manifolds 
 $X^{\rm (ref)}_{\rm free}$ and $X^{\rm (ref)}_{\rm tor}$ introduced above.
 By using it, the partition function on $X$ is defined for this $Y$.

Now let us return to the definition of the WZW term when $Y$ or $Y_0$ does not exist. 
Consider an $\h H$-bundle on $X$. By using the Goldstone field $U$, the topology of the $G$-bundle on $X$ is also determined by the $\h H$-bundle.

When $Y$ does not exist, the choice of the phase of the partition function just corresponds to the generalized $\theta$-angles of the $G$ gauge field $A$.
This $\theta$-angle is that of the background field, so it is not relevant for the dynamics of the Goldstone bosons. 
Therefore, we restrict our attention to the case that $Y$ exists.

The interesting case is that $Y$ exists but $Y_0$ does not. 
For example, it can happen that an $\h H$ bundle is topologically nontrivial but after embedding $\h H \to G$ the topology becomes trivial as a $G$ bundle. 
In this case, 
there is a generalized $\theta$-angle of the $\h H$-bundle and it can be regarded as a topological term of the Goldstone bosons. 

To make the discussion more explicit, let us consider the case of the $d=4$ QCD-like theories with 
gauge group $\SO(N_c)$ or $\Spin(N_c)$ as an example.\footnote{This problem is also investigated in \cite{Lee:2020ojw}. 
I thank Yuji Tachikawa for helpful discussions.} 
The Lie algebra of the gauge group is $\so(N_c)$.
As the matter content, we introduce $N_f$ flavors of massless Weyl fermions in
the fundamental ($N_c$-dimensional) representation of $\so(N_c)$. There is a global symmetry $G=\SU(N_f)$ which acts on the $N_f$ Weyl fermions.
For some range of $N_c$ and $N_f$, it is believed that this symmetry is spontaneously broken to $H=\SO(N_f)$ by strong dynamics,
and the low energy degrees of freedom is only the Goldstone bosons $G/H$, possibly also with some topological degrees of freedom which we discuss later. 

If the gauge group is $\SO(N_c)$ rather than $\Spin(N_c)$, 
we can introduce a discrete $\theta$-angle of the $\SO(N_c)$ gauge group~\cite{Aharony:2013hda}.
To see this, first notice that $\SO(N_c)$ has a subgroup $\SO(2) \subset \SO(N_c)$
and we can introduce magnetic fluxes $\sF$ of the $\SO(2)$ gauge field using the isomorphism $\SO(2) \cong \U(1)$. 
However, the difference between $\SO(2)$ and $\SO(N_c)$ is that
$\pi_1(\SO(2))=\bZ$ while $\pi_1(\SO(N_c)) = \bZ_2$ for $N_c \geq 3$. Thus, magnetic fluxes are topologically stable only mod $2$.
The topological classification of such fluxes is done by the cohomology group $H^2(X, \bZ_2)$,
and we define $\sF $ as an element of $ H^2(X, \bZ_2)$. Now the discrete $\theta$-angle is roughly defined as
\beq
\exp \left(  \pi \i \int_X \frac{1}{2} \sF \wedge \sF \right),
\eeq
where we have used differential form notation although $\sF$ is actually a cohomology element with $\bZ_2$ coefficients. 
More precisely, $\sF \wedge \sF$ is defined by Pontryagin square as an element of $H^2(X, \bZ_4)$.
On spin manifolds, $\int \sF \wedge \sF$ is known to be even.\footnote{
If we can embed $\sF$ in $\U(1) \cong \SO(2) \subset \SO(N_c)$, this statement can be easily shown by the Atiyah-Singer index theorem.
More generally, we have $\sF \wedge \sF \equiv \nu_2 \wedge \sF \mod 2$ where $\nu_2 \in H^2(X, \bZ_2)$ is the Wu class.
We have $\nu_2=0$ on spin manifolds. }
Thus the above discrete $\theta$-angle gives a sign factor $\pm 1$.

We denote the $\SO(N_c)$ group with and without the discrete $\theta$-angle as $\SO(N_c)_-$ and $\SO(N_c)_+$, respectively~\cite{Aharony:2013hda}.
Thus we have three gauge theories with the algebra $\so(N_c)$, 
\beq
\Spin(N_c), \qquad \SO(N_c)_+, \qquad \SO(N_c)_- . \label{eq:threegroup}
\eeq
We take the usual continuous $\theta$-angle associated to instantons to be zero.\footnote{
We take $N_c$ to be greater than $4$ to be in a generic situation.
In this case the discrete and continuous $\theta$-angles are completely independent parameters.} 
More precisely, we can make the continuous $\theta$-angle to be zero for massless quarks by axial rotations.
But we will also consider the situation in which we introduce quark masses, and in that case we only consider positive real mass parameters
so that there is no continuous $\theta$-angle. 

Now we want to determine the discrete $\theta$-angle of the low energy theory of Goldstone bosons
for a given UV group listed above. The system is strongly coupled and we can only guess the answer without a proof.

First consider the case $\SO(N_c)_+$ which does not have the discrete $\theta$-angle. This group has the following special feature.
The continuous part of the dynamics of the gauge field is believed to confine.
However, it is believed that the discrete part described by $\sF \in H^2(X, \bZ_2)$ remains 
as the low energy degrees of freedom. If we add masses to the quarks,
this $\sF$ is the sole low energy degrees of freedom.
In fact, $\SO(N_c)_+$ can be obtained
by starting from $\Spin(N_c)$ and gauging its 1-form center symmetry~\cite{Gaiotto:2014kfa}.
The 2-form gauge field for the 1-form center symmetry is $\sF$.
Therefore, if $\Spin(N_c)$ has a trivial gapped vacuum for the case of massive quarks,
the $\SO(N_c)_+$ theory has the 2-form $\bZ_2$ gauge field. See \cite{Aharony:2013hda,Tachikawa:2014mna,Gaiotto:2014kfa}
for more discussions. 

We assume that we can describe the low energy theory of the case $\SO(N_c)_+$ by taking the hidden local gauge group $\h H$ as 
$\h H = \Spin(N_f)$. This is a double cover of the unbroken subgroup $H = \SO(N_f)$ of the global symmetry $G = \SU(N_f)$. 
One of the reasons of this assumption is as follows. The Goldstone field $U$ spontaneously breaks the hidden local
gauge group as $\Spin(N_f) \to \bZ_2$. Thus there remains a 1-form $\bZ_2$ gauge field.
A 1-form $\bZ_2$ gauge field and a 2-form $\bZ_2$ gauge field are dual descriptions of each other.
Therefore, the low energy degrees of freedom matches.

Both the UV $\SO(N_c)_+$ and the IR $\h H = \Spin(N_f)$ have 1-form symmetries
and hence they can be coupled to a background 2-form field $\sG \in H^2(X, \bZ_2)$. 
The coupling of the UV theory to $\sG$ is simply given by 
\beq
\int_X \sG \wedge \sF, \label{eq:EMcoupling}
\eeq
where we have omitted a factor $\pi \i$.
On the other hand, the coupling of $\sG$ to the IR theory is done by requiring that
the $\bZ_2$ flux of the hidden local gauge group $\h H = \Spin(N_f)$ is given by $\sG$.

We have to be a little more precise about the coupling to $\sG$ in the low energy theory.
If $\frac{1}{2} \int_X  \sG \wedge \sG$ has a nontrivial value, this is the situation in which we cannot take the extension $Y_0$
of $X$. The reason is that if $Y_0$ exists, we get $\frac{1}{2} \int_X  \sG \wedge \sG = \frac{1}{2} \int_Y  \d (\sG \wedge \sG) =0$.
Therefore, there is some ambiguity in the definition of the WZW term corresponding to the non-existence of $Y_0$, and this 
leads to the discrete $\theta$-angle. 

An explicit example of configurations with a nontrivial value of $\frac{1}{2} \int_X  \sG \wedge \sG$  can be constructed as follows. We take $X = T^4 = T^2_1 \times T^2_2$,
and we introduce nonzero 2-form $\bZ_2$ fluxes on each of the $T^2$ factors $T_1^2$ and $T_2^2$. 
Namely we take $\sG = \sG_1 + \sG_2$, where $\sG_1 \in H^2(T^2_1,\bZ_2)$ and $\sG_2 \in H^2(T^2_2,\bZ_2)$
with $\int_{T^2_1} \sG_1 =\int_{T^2_2} \sG_2 = 1 \mod 2 $. Then we get $\frac{1}{2} \int_X  \sG \wedge \sG =1 \mod 2$.
We assume the generic case $N_f \geq 5$. Then take the Goldstone field $U$ to be 
\beq
U = U_1 \oplus U_2  \oplus 1 \in \SU(2) \times \SU(2) \times \SU(N_f-4) \subset \SU(N_f).  \label{eq:Uexplicit}
\eeq
Very explicitly, by taking coordinates $(x_1, x_2)$ of the $T^2_1$ with $x_1 \sim x_1 +1$ and $x_2 \sim x_2 + 1$, we take for $0 \leq x_1 \leq 1$,
\beq
U_1(x_1,x_2) =&\  (1 - x_1+ x_1\cos 2\pi x_2)\cdot 1 + ( \sqrt{x_1}\sin 2\pi x_2)\cdot \i \sigma^2 \nonumber \\
&+  \sqrt{x_1(1-x_1)  }(1-\cos 2\pi x_2)\cdot \i \sigma^1, \qquad 0 \leq x_1 \leq 1, \label{eq:twisted}
\eeq
where $\sigma^i~(i=1,2,3)$ are Pauli matrices. Notice that it is not periodic under $x_1 \to x_1+1$.
We have
\beq
U_1(x_1=0, x_2) & = 1, \nonumber\\
U_1(x_1=1, x_2) &= (\cos 2\pi x_2)\cdot 1   + (\sin 2\pi x_2)\cdot \i \sigma^2 := V(x_2) \in \SO(2).
\eeq
This means that the identification of $x_1 =0$ and $ x_1=1$ is done by using $V(x_2)$ as
a transition function of the hidden local gauge group $\h H $. This nontrivial transition function $V(x_2)$
gives the nontrivial flux $\sG_1$ on $T^2_1$ since the function $V(x_2)$ gives an element of $\pi_1(\SO(2))$
which is still nontrivial in $\pi_1(\SO(N_f))$. 
We also take $U_2$ to be the same functional form with $(x_1,x_2)$
replaced by the coordinates $(x_3,x_4)$ of the $T^2_2$ in $X = T^2_1 \times T^2_2$.

The WZW term $\Phi_{\rm WZW}(X)$ in the above configuration is either $\pm 1$, at least if $N_f \geq 5$.
To see this, notice that 
\beq
\sigma_1 U_1(x_1, x_2) \sigma_1 = U_1(x_1, -x_2). \label{eq:parity}
\eeq
The matrix $\sigma_1$ is not an element of $\SO(2)$. But it can be embedded in $\SO(3)$ as $\sigma_1 \oplus (-1)$.
This embedding is possible by using $N_f - 4 \geq 1$ components which have played no role in the above construction. 
Then, \eqref{eq:parity} implies that $U_1(x_1, x_2)$ is invariant under the orientation reversal $x_2 \to -x_2$ up to
$\SO(3)$ symmetry transformations. Therefore we get $\Phi_{\rm WZW}(\o X) = \Phi_{\rm WZW}(X)$ and hence
\beq
(\Phi_{\rm WZW}(X))^2 = \Phi_{\rm WZW}(X)\Phi_{\rm WZW}(\o X) =  \Phi_{\rm WZW}(X)\Phi_{\rm WZW}( X)^* = 1.
\eeq
This can also be shown by taking a 3-dimensional manifold $W$ whose boundary is two copies of $T^2_1$,
and set $Y_0 = W \times T^2_2$. Such $W$ can be constructed from \eqref{eq:parity}.
Notice that the WZW term of $U$ is just a sum of WZW terms of $U_1$ and $U_2$.
Also notice that $U_1$ only depends on the coordinates of $W$, while $U_2$ only depends on $T^2_2$,
so each of them is trivial on $W \times T^2_2$.

There are two possibilities $\Phi_{\rm WZW}(X) = \pm 1$ for the above configuration \eqref{eq:Uexplicit}.
We propose that if the UV theory does not have a discrete $\theta$-term $\exp(\pi \i \cdot \frac{1}{2} \int_X  \sG \wedge \sG)$ for the background field $\sG$,
then the low energy theory is such that $\Phi_{\rm WZW}(X) = +1$.
This is an assumption about the strong dynamics, but the intuition behind it is as follows. 
The flux $\sG$ can be regarded as the electric flux of the UV gauge group $\SO(N_c)$~\cite{Witten:1983tw}.
In fact, $\sF$ is the magnetic flux of the UV gauge group, and \eqref{eq:EMcoupling} implies
that $\sG$ and $\sF$ are ``electromagnetic dual'' of each other.
There is no particular reason that electric fluxes of the UV gauge field should produce a nontrivial phase factor.
For example, we can add masses to the quarks so that the UV theory goes to a pure Yang-Mills without a $\theta$-angle.
It is hard to imagine that some nontrivial phases are produced in such a pure Yang-Mills. 
By this intuition, we propose $\Phi_{\rm WZW}(X)=+1$ when the UV theory does not contain an explicit $\theta$-term $\frac{1}{2} \int_X  \sG \wedge \sG$.

We remark that it is sufficient to fix $\Phi_{\rm WZW}(X) $ on the above $X = T^2_1 \times T^2_2$
to give a complete specification of the $\theta$-angle up to the $\theta$-angles of the background fields.
This fact follows from the result for the bordism group $\Omega^{\spin \times \SO(N_f) }_4 \cong \bZ \oplus \bZ \oplus \bZ_2$
as computed in \cite{Garcia-Etxebarria:2018ajm,Lee:2020ojw}. The free part $ \bZ \oplus \bZ$ is related to the continuous
$\theta$-angles of gravity and $G=\SU(N_f)$ under the embedding $H=\SO(N_f) \to \SU(N_f)=G$.  
The torsion part $\bZ_2$ is related to the discrete $\theta$-angle relevant to the sign of $\Phi_{\rm WZW}$.
The torsion flux becomes trivial when $\SO(N_f)$ is embedded in $\SU(N_f)$ since $\pi_1(\SU(N_f)) =0$. 

It is now easy to determine the low energy $\theta$-angle when the UV gauge group is $\Spin(N_c)$ or $\SO(N_c)_-$.
The group $\Spin(N_c)$ is obtained by making $\sG$ dynamical without the $\theta$-term $\frac{1}{2} \int_X  \sG \wedge \sG$.
By integrating over $\sG$, the coupling \eqref{eq:EMcoupling} forces $\sF=0$, which is just $\Spin(N_c)$.
On the other hand, $\SO(N_c)_-$ is obtained by making $\sG$ dynamical with the $\theta$-term $\frac{1}{2} \int_X  \sG \wedge \sG$ included.
By integrating over $\sG$, we get
\beq
\int_X \sG \wedge \sF + \frac{1}{2} \sG \wedge \sG \xrightarrow{\text{integrating }\sG} \int_X \frac{1}{2} \sF \wedge \sF.
\eeq
This is the discrete $\theta$-angle of the UV gauge group and hence it is $\SO(N_c)_-$.
In both cases, the IR hidden local gauge group is now $\h H = \SO(N_f)$.
The WZW term is $\Phi_{\rm WZW}(X) = + 1$ if the UV group is $\Spin(N_c)$, and it is $\Phi_{\rm WZW}(X) = - 1$
if the UV group is $\SO(N_c)_-$. We denote the hidden local gauge groups corresponding to these WZW terms as $\SO(N_f)_+$
and $\SO(N_f)_-$, respectively. 

We can summarize what we have discussed above as follows.
In the UV, we have the gauge group which is either $\SO(N_c)_{\pm }$ or $\Spin(N_c)$.
In the IR, we have the hidden local gauge group which is either $\SO(N_f)_{\pm }$ or $\Spin(N_f)$
in the sense described in the previous paragraph.
We propose the following matching between them:
\beq
\begin{array}{c|c|c|c}
\text{UV gauge} ~ & ~\Spin(N_c) ~& ~\SO(N_c)_{ + } ~&~ \SO(N_c)_{-} ~  \\  \hline 
\text{IR} ~\h H~&~\SO(N_f)_+ ~&~ \Spin(N_f) ~&~ \SO(N_f)_{-} ~
\end{array}. \label{eq:UVIR}
\eeq
Let us briefly recapitulate the discussions.
In the case of the UV group $\SO(N_c)_{ + } $, the $\bZ_2$ magnetic flux $\sF$ of $\SO(N_c)_{ + } $
is summed over which remains as low energy degrees of freedom,
while the IR theory has the $\bZ_2$ gauge field coming from the spontaneous symmetry breaking 
$\h H =\Spin(N_f) \to \bZ_2$.
These degrees of freedom are dual to each other. 
There is no twisted configuration of the Goldstone field like \eqref{eq:twisted} since $\h H$ is simply connected.
In the case of the UV group $\Spin(N_c)$, the flux $\sF$ is set to zero, 
and correspondingly the IR group $\h H$ is completely higgsed.
In the case of the UV group $\SO(N_c)_{-}$, the flux $\sF$ is summed over with the additional theta term $\frac12 \int \sF \wedge \sF$.
In the IR, this additional theta term leads to the nontrivial values of the WZW term in configurations like \eqref{eq:twisted}.

Let us make some comments on the above identification.\footnote{I would like to thank Y.~Tachikawa for pointing out an error in this paragraph in the first version of the paper.}
In QCD-like theories, the hidden local gauge group may be interpreted as a gauge group for $\rho$-mesons~\cite{Bando:1984ej}.
Moreover, there may be some analogy~\cite{Komargodski:2010mc} between $\rho$-mesons as hidden local gauge fields and 
Seiberg dual gauge fields in supersymmetric QCD.
If we interpret $\rho$-mesons as gauge fields for $\h H$, the above correspondence \eqref{eq:UVIR}
means that the UV group and $\h H$ are electromagnetic duals of each other.
A subtle point in supersymmetric QCD is that the Seiberg duality for $\so(N_c)$ gauge theories~\cite{Seiberg:1994pq,Intriligator:1995id}
is better regarded as an ``electro-dyonic duality''. 
The reason is that confinement in the original theory is dual to condensation of some dyonic particle in the dual theory
rather than dual quarks~\cite{Intriligator:1995id}.
The identifications of the groups $\SO(N_c)_+  \leftrightarrow \SO(N_f -N_c +4)_+$, $\Spin(N_c)  \leftrightarrow \SO(N_f -N_c +4)_-$
and $\SO(N_c)_-  \leftrightarrow \Spin(N_f -N_c +4)$ shown in \cite{Aharony:2013hda} imply that the Seiberg duality is an electro-dyonic duality. 
In our case, the correspondence $\SO(N_c)_+ \leftrightarrow \Spin(N_f)$ is unavoidable just by 
the counting of the low energy degrees of freedom. Thus it seems that the correspondence in non-supersymmetric QCD
is an electromagnetic duality.\footnote{In $\cN=2$ supersymmetric $\so(3) \cong \su(2)$ pure Yang-Mills, we know that there is a magnetic point and a dyonic point~\cite{Seiberg:1994rs}.
It seems that the magnetic point is relevant for the low energy phase of non-supersymmetric QCD without a $\theta$-angle, 
while the dyonic point generalizes to the $\cN=1$ Seiberg duality. } 
As a check, we remark that a $\bZ_2$ gauge theory with a 2-form gauge field $\sF$
which is obtained after confinement of $\SO(N_c)$ is an invertible field theory (i.e. no degrees of freedom) if there is a topological term $\frac{1}{2} \sF \wedge \sF$
(see \cite{Bhardwaj:2020ymp} for a recent discussion).
Then one can see that the low energy degrees of freedom is consistent.

In any case, the main message is that the IR $\theta$-angle of Goldstone bosons depends on the UV theory. 
This message does not depend on whether our identifications are correct or not. 
If the identifications of the UV gauge groups and the values $\Phi_{\rm WZW}= \pm 1$ for \eqref{eq:Uexplicit}
are opposite to the ones proposed above, we just need to exchange $\SO(N_f)_+$ and $\SO(N_f)_-$
in the table \eqref{eq:UVIR}. 
We remark that all these theories have the same anomaly under $G = \SU(N_f)$,
so the choice of the $\theta$-angle has nothing to do with the anomaly, as the general argument in the previous section clearly shows. 

\acknowledgments
I would like to thank Yuji Tachikawa for helpful discussions.
KY is in part supported by JSPS KAKENHI Grant-in-Aid (Wakate-B), No.17K14265.


\bibliographystyle{JHEP}
\bibliography{ref}

\end{document}